\documentclass[aps,prl,twocolumn,floatfix]{revtex4}
\usepackage{amssymb}

\usepackage{epsfig}
\usepackage{amsmath}

\begin{document}

\title{The Fermi edge singularity of spin polarized electrons}
\author{P. \surname{Plochocka-Polack}$^{1}$}
\email{Paulina.Plochocka@grenoble.cnrs.fr}
\author{J. G. \surname{Groshaus}$^{2}$}
\author{M. \surname{Rappaport}$^{1}$}
\author{V. \surname{Umansky}$^{1}$}
\author{Y. \surname{Gallais}$^{2}$}
\author{A. \surname{Pinczuk}$^{2}$}
\author{I. \surname{Bar-Joseph}$^{1}$}
\affiliation{$^{1}$Department of Condensed Matter Physics, The Weizmann Institute of
Science, Rehovot, Israel\\
$^{2}$Department of of Physics and of Appl. Physics and Appl. Mathematics,
Columbia University, New York, NY 10027}
\date{\today }

\begin{abstract}
We study the absorption spectrum of a two-dimensional electron gas (2DEG) in
a magnetic field. We find that that at low temperatures, when the 2DEG is
spin polarized, the absorption spectra, which correspond to the creation of
spin up or spin down electron, differ in magnitude, linewidth and filling
factor dependence. We show that these differences can be explained as
resulting from creation of a Mahan exciton in one case, and of a power law
Fermi edge singularity in the other.
\end{abstract}

\maketitle

%% INTRODUCTION

The role of Coulomb interactions in the optical absorption spectrum of a
Fermi sea of electrons has attracted interest for several decades. The
response of the electron gas to the sudden creation of the hole attractive
potential is manifested in a singularity at the photon energy for which an
electron is excited to the Fermi level. This singularity, known as the Fermi
edge singularity (FES), was first studied by Mahan~\cite{Mahan67a,Mahan67b}
in the context of the X-ray absorption edge of metals and the interband
spectrum of degenerate semiconductors. Using a ladder diagram approach he
was able to show that the absorption is singular at the threshold energy
%TCIMACRO{\U{127}}%
%BeginExpansion
h{\hskip-.2em}\llap{\protect\rule[1.1ex]{.325em}{.1ex}}{\hskip.2em}%
%EndExpansion
$\omega =E_{G}+p_{F}^{2}$ $/2\mu $, where $E_{G}$\ is the gap energy, $p_{F}$
is the electron Fermi momentum, and $\mu $\ is the electron-hole reduced
mass. At higher energies, the absorption intensity decays as $(\omega -$ $%
\omega _{0})^{-\alpha }$, where $\alpha $ is a dimensionless coupling
constant describing the interaction between the electrons and the deep hole
created in the absorption process. An exact expression for the absorption
spectrum was later derived by Combescot and Noziers~\cite%
{Nozieres69,Combescot71}, and practically verified Mahan's predictions.

A two-dimensional electron gas (2DEG) in modulation doped quantum wells has
proven to be particularly convenient for observing and studying the FES. The
ability to vary the electron density, $n_{e}$, and the ratio between
temperature and Fermi energy provides an important handle on the behavior of
the singularity. Indeed, in the last two decades there has been a rich body
of theoretical~\cite{Ruckenstein87,Sham90,Hawrylak91,Westfahl98} and
experimental work~\cite%
{Skolnick87,Kalt87,Lee87,Brown96,Huard00,Yusa00,Kim92,Bar-Ad94,Brener95,Chen92,Rubio97,Geim94}
on various aspects of the singularity in this system. In particular,
photoluminescence measurements have shown that the singularity persists when
a magnetic field is applied \cite{Skolnick87,Chen92,Rubio97}: the envelope
of the Landau levels (LL) resembles the FES lineshape at zero magnetic field%
\textit{~}~\cite{Westfahl98}. The role of the spin degree of freedom was,
however, commonly assumed to be limited to the degeneracy in the density of
states. Yet, it is plausible to expect that the 2DEG spin polarization would
affect the FES in a profound way: electron-electron scattering, which plays
a key role in the singularity \cite{Mahan67b}, is strongly suppressed when
the spin of the photo-excited electron is opposite to that of the Fermi sea
electrons.

In this paper we study the absorption spectrum of a high mobility 2DEG in a
magnetic field. Our main finding is that at low temperatures, when the 2DEG
is spin polarized, the absorption spectra which correspond to the creation
of\ a $|\downarrow >$ or $|\uparrow >$ electron differ in magnitude,
linewidth and filling factor dependence. We show that these differences can
be explained as resulting from the creation of a Mahan exciton \cite%
{Mahan67a} in one case, and of a power law FES \cite{Mahan67b} in the other.

Measuring the absorption spectrum of a single GaAs quantum well (QW) at low
temperatures has always been a difficult task. Here we introduce a new
technique for measuring the absorption spectrum of a single QW, using very
weak white light in a reflection geometry. The idea is growing a cavity
structure in which the QW is located at an anti-node of a standing wave
formed by the optical field. The structure consists of a Bragg reflector,
made of $20$ pairs of $\lambda /4$ layers of AlAs and Al$_{0.3}$Ga$_{0.70}$%
As ($\lambda =800$ nm), with its top layer is at a distance of $3\lambda /2$
from the surface. The Bragg reflector and the sample surface form a
microcavity with a broad stop-band, in which all wavelengths are back
reflected ($R\sim 1$). A modulation-doped GaAs QW, which is 20 nm wide and
is embedded in Al$_{0.38}$Ga$_{0.62}$As barriers, is grown in the middle of
the cavity, at a distance of $3\lambda /4$ from the Bragg mirror and the
surface. Figure 1 shows the calculated reflectivity spectra, assuming
typical excitonic parameters of a GaAs QW. The broad stop-band is seen to
extend from $1.47$ to $1.63$ eV, with a deep notch at the exciton energy , $%
\sim 1.538$ eV. We find that the exciton lineshape is very sensitive to the
location of the QW within the cavity, and one can get a Lorentzian or
dispersive lineshape by varying it~\cite{Zheng88}. The inset shows a
reflectivity measurement of an optimized sample with a single
modulation-doped QW at $B=9$ T. It can be seen that the reflectivity at the
lowest LL energy is $\sim 0.3$, corresponding to an order of magnitude
enhancement relative to the single pass absorption. The whole structure is
grown on an $n^{+}$ layer that serves as a back gate. The wafer was
processed to form a mesa structure, with ohmic contacts to the QW and to the
back gate, such that $n_{e}$ can be tuned in the range $(0.4-3)\times
10^{11} $ cm$^{-2}$. The measured electron mobility is $\sim 1\times 10^{6}$
cm$^{2}$ V$^{-1}$ s$^{-1}$. We used a tungsten halogen lamp to illuminate
the sample, and selected a spectral range of 20~nm, centered around the
heavy hole (hh) transition of the QW. The reflected light is analyzed with a
circular polarizer, then dispersed in a spectrometer and detected in a CCD
camera. This configuration allowed us to measure the absorption spectrum
over a broad spectral range in a fraction of a second using sub-nW light
intensity.

\begin{figure}[t]
\begin{center}
\includegraphics[width=.4\textwidth]{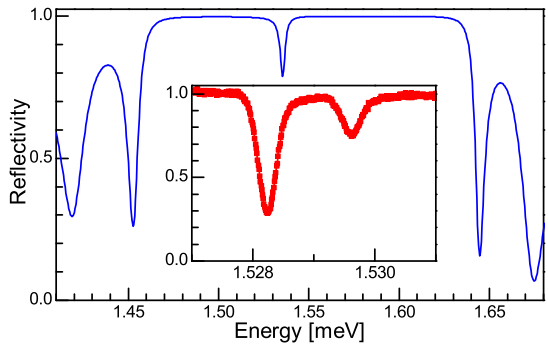}
\end{center}
\caption{The calculated reflectivity of the Bragg mirror structure with a
GaAs QW placed in the anti-node of the electric field. The inset shows the
measured reflectivity of a single modulation doped quantum well at $B=9$ T.
The two peaks correspond to transitions from the heavy and light hole bands
to the lowest LL.}
\label{fig1}
\end{figure}

The experiment is performed in two systems: (i) A dilution refrigerator with
optical windows at its base temperature of $40$ mK. (ii) A He$^{_{4}}$
cryostat with a fiber based system for illumination and collection and a
temperature range of $1.8-4.2$K. The magnetic field in both systems is
applied parallel to the the growth direction. Changing the sign of $B$ while
keeping the circular polarizer fixed corresponds to changing the circular
polarization of the light: at positive fields we detect transitions from the
hh band to the lower electron Zeeman spin subband (LZ) and at negative field
- to the upper electron Zeeman spin subband (UZ). We label the transition to
LZ as $\sigma ^{+}$, and to UZ as $\sigma ^{-}$.

\begin{figure*}[tb]
\begin{center}
\includegraphics[width=.8\textwidth]{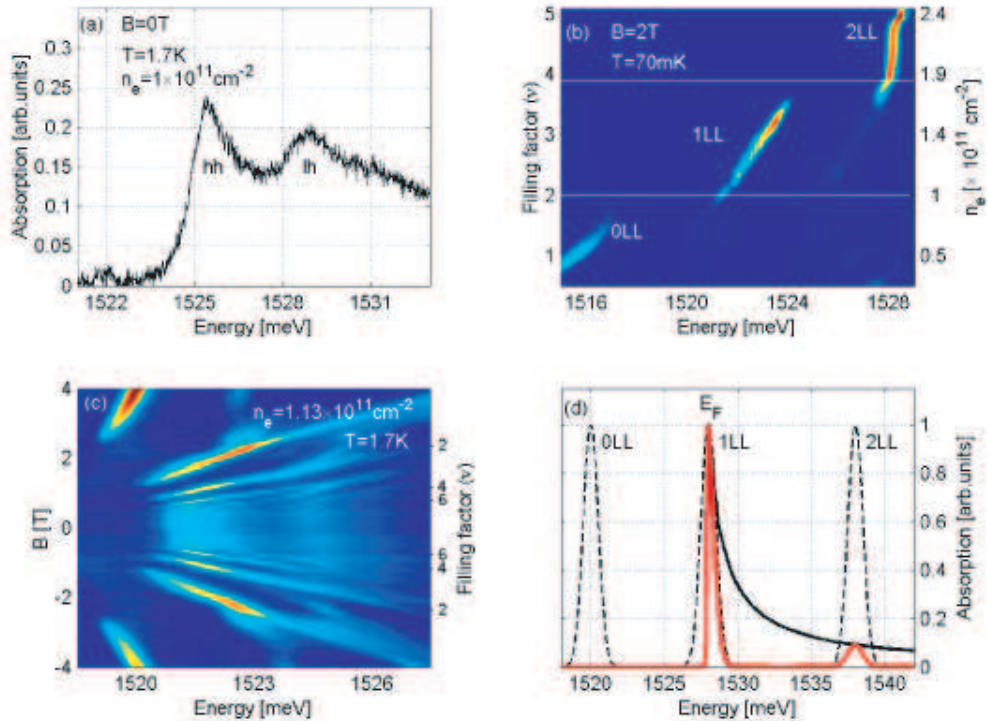}
\end{center}
\caption{(a) Absorption spectrum at $B=0$, exhibiting the characteristic FES
lineshape. (b) The $\protect\sigma ^{-}$\ absorption spectra as a function
of $n_{e}$ in the range ($0.4-2.4)\times 10^{11}$ cm$^{-2}$ (the shift of
the absorption lines to higher energies is due to the gate electric field).
(c) The absorption spectra as a function of magnetic field. The absorption
strength in (b) and (c) is color coded such that red and blue correspond to
high and low absorption, respectively. (d) The single particle joint density
of states (dashed black line), the FES function (solid black line) and the
resulting absorption (red line) for the case where $E_{F}$ is at the center
of the 1LL. }
\label{fig2}
\end{figure*}

Figure 2a shows a typical absorption spectrum at $B=0$ and $n_{e}=1\times
10^{11}$ cm$^{-2}$. Two peaks, corresponding to transitions from the heavy
and light hole bands, can be clearly resolved. We note that the peaks are
broad and have the characteristic asymmetric FES lineshape. Indeed, at this
electron density the exciton and trion, which characterize the spectrum at
lower densities, no longer exist. As we turn on the magnetic field the broad
FES line splits into discrete LLs. Figure 2b shows a compilation of
absorption spectra at $B=2$ T, in the range $4\times 10^{10}<n_{e}<2.4\times
10^{11}$ cm$^{-2}$, which corresponds to the filling factor range $1\lesssim
\nu <5$. It can be seen that the absorption into a certain LL is strong only
within a limited range. For example, absorption into the first LL (1LL) is
significant only in the range $2<\nu <4$: at electron densities above and
below this range the absorption is substantially weaker. Similar behavior is
observed when we fix $n_{e}$ and change $B$. Figure~\ref{fig2}c shows a
compilation of absorption spectra measured at $n_{e}=1.13\times 10^{11}$ cm$%
^{-2}$ for both signs of the magnetic field. For example, it can be seen
that the 1LL absorption at negative fields is turned on at $B\simeq -1$ T ($%
\nu =4$), reaches a maximum at $B\simeq -2$ T, and then gradually decays
with increasing field. The appearance of absorption to a certain LL above a
critical field (or below a critical $n_{e}$) could be explained by a band
filling argument: Absorption to a LL is inhibited as long as it is full, and
becomes allowed when its degeneracy is increased by increasing $B$ (or its
occupation is decreased by decreasing $n_{e}$) \cite{Groshaus2004}. The
behavior at the other limit is surprising: Why would the absorption to an
empty LL be lower than to a partially filled one?

The answer to this question lies in the FES behavior, which is schematically
depicted in Fig. 2d. The dashed line describes the LL single particle joint
density of states, while the solid black line depicts the FES function when
the Fermi energy ($E_{F}$) is at the center of the 1LL. The resulting
absorption is described by the red line, which is the product of the dashed
and solid black lines. It can be seen that the absorption of the 2LL is
weak, even though it is empty of electrons. This description implies that
the absorption is strong only at the LL at which $E_{F}$ resides, as indeed
is observed in the experiment.

\begin{figure}[t]
\begin{center}
\includegraphics[width=.4\textwidth]{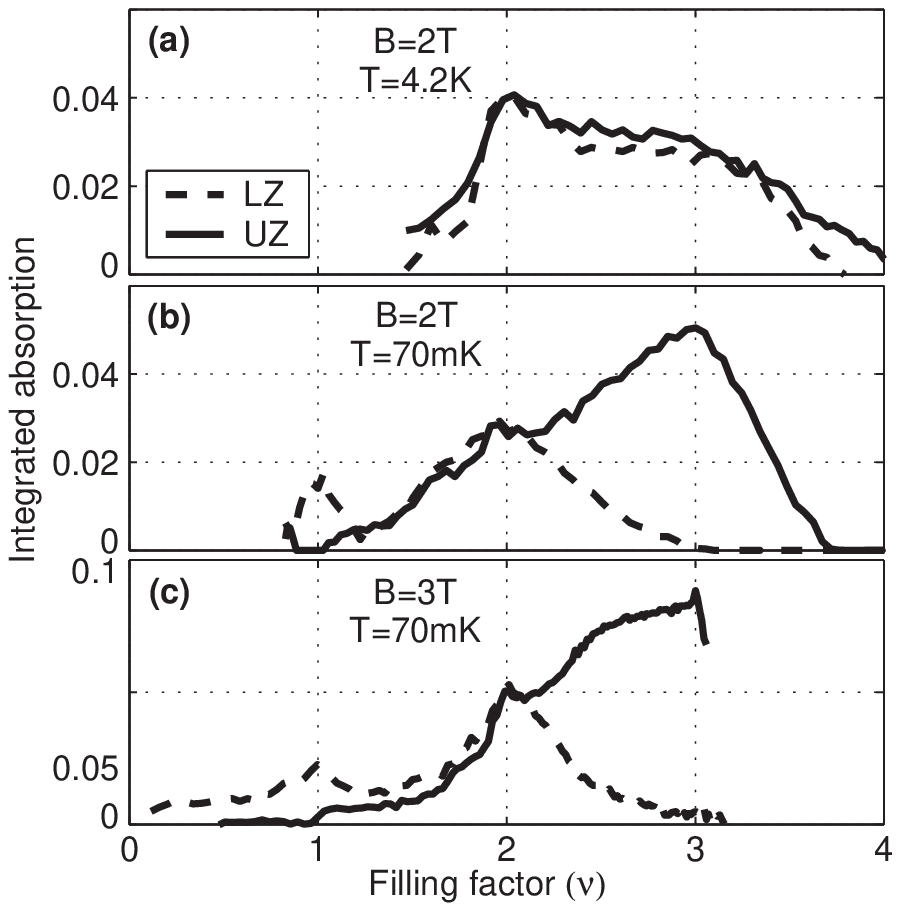}
\end{center}
\caption{Integrated absorption to the UZ and LZ of the 1LL as a function of
filling factor, while keeping the magnetic field constant: (a) $4.2$ K and $%
2 $ T, (b) $70$ mK and $2$ T, and (c) $70$ mK$\ $and $3$T.}
\label{fig3}
\end{figure}

Let us now consider the role of the electron spin polarization in the FES.
We start with measurements done at a high temperature of $4.2$ K and weak
magnetic field of $2$ T, where the electrons should be spin depolarized.
Figure 3a shows the integrated absorption at the 1LL as a function of
filling factor, while keeping $B$ constant. Considering the scheme of Fig.
2d, this curve describes the dependence of the area under the central peak
on $n_{e}$. The onset of absorption at $\nu =4$ and its drop at $\nu <2$ are
clearly visible. As expected for spin depolarized 2DEG, the absorption to
both Zeeman levels is nearly identical.

This behavior drastically changes as we cool down the sample and the
electrons become spin polarized. Figure 3b compares the integrated
absorption to the UZ and LZ at 70 mK and $B=2$ T. We can see that at $\nu <2$%
, where $E_{F}$ is in the 0LL and both Zeeman levels of the 1LL are empty,
the absorption is essentially the same (there is a difference near $\nu =1$%
). At $\nu >2$ the two absorption lines behave markedly different: The
absorption to the LZ \textit{decreases} monotonically and vanishes at $\nu
=3 $, when this level is fully occupied and $E_{F}$ jumps to the UZ. The
absorption into the UZ, on the other hand, \textit{increases} first, reaches
a maximum at $\nu =3$, \ and then falls steeply and vanishes slightly below $%
\nu \simeq 4$. In fact, in the range $2<\nu <3$ the sum of the two
absorption curves is nearly constant: The drop in the absorption of the LZ
line is compensated by the rise in the absorption of the UZ. We can see that
this picture repeats when we increase the field to $3$ T (Fig. 3c): The
increase in degeneracy with increasing the field causes the UZ and LZ
absorption to increase by $\sim 50\%$, yet the dependence on $\nu $ remains
the same. Furthermore, we find the same qualitative behavior at the 2LL at
the corresponding $\nu $'s.

\begin{figure}[t]
\begin{center}
\includegraphics[width=.4\textwidth]{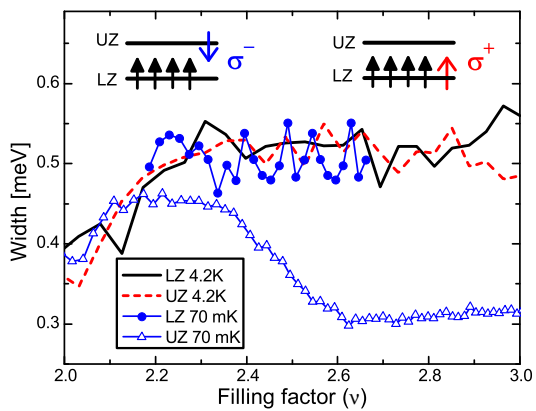}
\end{center}
\caption{Filling factor dependence of the UZ and LZ linewidth (full width at
half maximum) at $4.2$ K and $70$ mK. The behavior of UZ linewidth at $70$
mK is markedly different. The upper schemes describe the spin polarizations
of the photo-electron and the Fermi sea for $\protect\sigma ^{-}$ and $%
\protect\sigma ^{+}$ excitation at $2<\protect\nu <3$.}
\label{fig4}
\end{figure}

The fact that the two absorption curves are identical at $4.2$ K implies
that the oscillator strength of the two Zeeman levels is the same, and the
observed differences at low temperature are due to the difference in
occupation of the two Zeeman levels. To explain this behavior we focus on
the range $2<\nu <3$. In this range absorption of a $\sigma ^{-}$ photon
into the UZ creates an electron with $S_{Z}=+1/2$ on a background of a Fermi
sea of $S_{Z}=-1/2$ electrons (left scheme, Fig. 4). On the other hand, when
a $\sigma ^{+}$ photon is absorbed the photo-created electron has the same
spin as the background 2DEG (right scheme, Fig. 4). Hence, the two
measurements represent different relations between the spin orientation of
the photo-excited and the background electrons, one of counter-polarization (%
$\sigma ^{-}$) and one of co-polarization ($\sigma ^{+}$).

In his first paper on the subject, Mahan~\cite{Mahan67a} considered the
screening of the electron-hole interaction by the Fermi sea of electrons and
predicted that an \textit{excitonic} state is formed in the absorption
process, with a binding energy $E_{B}=(2p_{F}^{2}$ /$\mu )\exp (-1/\Delta )$%
, where $\Delta =(1/2\pi p_{F}a_{B})\ln (1+4p_{F}^{2}/k_{s}^{2})$. This
so-called ''Mahan exciton'' was predicted to appear in the optical
absorption spectrum at an energy
%TCIMACRO{\U{127}}%
%BeginExpansion
h{\hskip-.2em}\llap{\protect\rule[1.1ex]{.325em}{.1ex}}{\hskip.2em}%
%EndExpansion
$\omega =E_{G}+p_{F}^{2}$ $/2\mu -E_{B}$. The calculation included the sum
of ladder diagrams representing the scattering of the electron by the
(screened) hole potential. In a subsequent paper~\cite{Mahan67b}, published
shortly after, the analysis was extended to include crossed-diagrams, which
describe processes in which the electron hole interaction is mediated by
electron-electron scattering. It was found that the bound state no longer
exists. Rather, the absorption spectrum has a power law singularity at the
Fermi energy. The notion of the Mahan exciton was therefore widely
considered as erroneous, ''resulting from an incomplete analysis. We argue
that Mahan's first analysis ~\cite{Mahan67a} is applicable when the
photo-excited electron spin is opposite to the 2DEG polarization. The
crossed-diagrams, which describe virtual processes in which the optically
excited electron is exchanged with an electron from the 2DEG having the%
\textit{\ same }spin, are forbidden. In the second case of co-polarization,
the crossed-diagrams are allowed and should be included, and we get back to
the classical description of the FES. We therefore suggest that in a $\sigma
^{-}$ absorption we create a Mahan exciton, which has a stronger oscillator
strength than the FES.

Experimental support for this argument comes from analysis of the width of
the absorption lines in the two processes. Figure 4 compares the full width
at half maximum of the $\sigma ^{-}$ and $\sigma ^{+}$ lines in the filling
factor range $2<\nu <3$ at high and low temperature. We find that at $4.2$ K
the width of the two lines is identical: Both exhibit a slight rise of the
width with increasing $\nu $, and then remain constant at a value of $0.5$
meV. The behavior at $70$ mK is radically different: While the width of the $%
\sigma ^{+}$ line follows the same behavior as at $4.2$ K, the width of the $%
\sigma ^{-}$ line drops to $0.3$ meV, almost half of the $\sigma ^{+}$
width. The fact that this narrowing of the $\sigma ^{-}$ peak occurs only at
low temperatures, oppositely to the behavior of the $\sigma ^{+}$ peak,
proves that it is related to the electron's spin polarization, rather than
to an intrinsic difference between the properties of the UZ and LZ levels.
The accumulated evidence, of a narrower and stronger $\sigma ^{-}$ peak, are
consistent with an excitonic resonance.

This work was supported by the Binational Science Foundation. We wish to
acknowledge Wiktor Maslana for his help in obtaining the data of Fig. 1, and
Alexander Finkel$^{\prime }$stein for fruitful discussions. The work at
Columbia University is supported the National Science Foundation under Award
Number DMR-03-52738, the Department of Energy under award DE-AIO2-04ER46133,
the Nanoscale Science and Engineering Initiative of the National Science
Foundation under NSF Award Numbers CHE-0117752 and CHE-0641523, and by the
New York State Office of Science, Technology and Academic Research (NYSTAR),
and by a research grant of the W.M. Keck Foundation.

\end{document}